\documentclass[a4paper,twocolumn,
secnumarabic, aps, superscriptaddress,
amssymb, amsmath, nobibnotes, showpacs]{revtex4}

\usepackage{graphicx}
\usepackage{dcolumn}

\begin{document}
\title{How proteins squeeze through polymer networks: a Cartesian lattice study}
\author{Annika Wedemeier}
\affiliation{BIOMS Center for Modelling and Simulation in the Biosciences,
 D-69120 Heidelberg, Germany}
\affiliation{Deutsches Krebsforschungszentrum, D-69120 Heidelberg,
Germany}
\author{Holger Merlitz}
\affiliation{Softmatter Lab, Department of Physics, Xiamen University,
Xiamen 361005, P.R.\ China}
\author{Chen-Xu Wu}
\affiliation{Softmatter Lab, Department of Physics, Xiamen University,
Xiamen 361005, P.R.\ China}
\author{J\"org Langowski}
\affiliation{Deutsches Krebsforschungszentrum, D-69120 Heidelberg,
Germany}

\date{\today}

\begin{abstract}
In this paper a lattice model for the diffusional transport of particles in
the interphase cell nucleus is proposed. The dynamical behaviour of single
chains on the lattice is investigated and Rouse scaling is verified.
Dynamical dense networks are created by a combined version of the bond
fluctuation method and a Metropolis Monte Carlo algorithm. Semidilute
behaviour of the dense chain networks is shown. By comparing diffusion of
particles in a static and a dynamical chain network , we demonstrate that
chain diffusion does not alter the diffusion process of small
particles. However, we prove that a dynamical network facilitates the
transport of large particles. By weighting the mean square displacement
trajectories of particles in the static chain network data from the dynamical
network can be reconstructed. Additionally, it is shown that subdiffusive
behaviour of particles on short time scales results from trapping processes in
the crowded environment of the chain network. In the presented model a protein
with 30 $nm$ diameter has an effective diffusion coefficient of $1.24\cdot
10^{-11}$ $m^2/s$ in a chromatin fiber network.
\end{abstract}


\maketitle

\section{Introduction} To accomodate the genome within the 
the cell nucleus, DNA is organized into chromatin. 
$5\%-12\%$ of the interphase cell nucleus are filled with a dense
network of chromatin fibers. In recent years, direct visualization of chromatin
structure and dynamics in live cells has been achieved through significant
advances in the field of microscopy and molecular biology \cite{levi}. The
growing awareness of thermal fluctuations and their impact on the chromatin
dynamics has modified the formerly common picture of an interphase nucleus
which contains randomly arranged, but static chromatin.

In \cite{marshall}
specific chromosome sites were tagged in living cells of \emph{Saccharomyces
cerevisiae} with green fluorescent protein and the motion of interphase
chromatin was measured at high resolution and in three dimensions. In
\cite{vazquez} chromosome motion was tracked in \emph{Drosophila spermatocyte
nuclei} by 3D fluorescence microscopy. A highly dynamic chromosome
organization was found governed by two types of motion: a fast, short range
component over a 1-2 $s$ time scale and a slower component related to long
range chromosome motion within the nucleus. The motion patterns are consistent
with random walks. These reports and others \cite{chubb} reveal that the motion
of specific chromatin regions is Brownian.  However, this motion is
limited to a nuclear subregion, i.e. a given chromatin segment is free to move
within only a limited subregion of the nucleus. 

Diffusive processes in the
cell play a key role in keeping the organism alive \cite{riggs, richter}:
Molecules transported through cell membranes, drugs on their way to respective
protein receptors, and proteins interacting with specific DNA sequences
constituting all of the biological functions of DNA \cite{berg, ptashne} are
diffusion controlled reactions. Furthermore, proteins approaching their
specific target sites on DNA are transported by diffusion or even facilitated
diffusion. However, the diffusional transport of molecules in the interphase
cell nucleus is fundamentally different from the normal kind of diffusion
which a molecule undergoes in a homogeneous fluid where the mean square
displacement of a molecule behaves linear in time $t$, $\langle
r^2(t)\rangle=6Dt$ with $D$ as the diffusion coefficient.  In the cell nucleus
the motion of proteins is strongly influenced by the presence of the "sticky
tangle" of chromatin fibers due to steric obstruction and thermal fluctuations
of the fibers. Fluorescence correlation spectroscopy (FCS) studies have shown
obstructed diffusion of autofluorescent proteins \cite{wachsmuth,
banks}. Obstructed diffusion or subdiffusion is characterized by $\langle
r^2(t)\rangle \sim t^\alpha$ with the anomalous diffusion exponent
$\alpha<1$. Other FCS measurements indicate that most of the nuclear space is
accessible to medium sized proteins by simple diffusion and that there is no
preference for interchromosome territory channels \cite{weidemann}. For small
proteins up to the size of eGFP-tetramers the entire intranuclear chromatin
network is freely accessible \cite{dross}.  It is still a matter of extensive discussion 
to what extent macromolecular mobility is affected by structural components of the
nucleus. This question was adressed in our recent work 
\cite{wedemeier1,wedemeier2} where we investigated the diffusional transport of
nonbinding and binding proteins in static polymer chain networks. However, it is known
that nuclear constituents much larger than the average mesh size of the chromatin
network can move in a random walk fashion \cite{goerisch}. This implies that the fluctuations of the 
network create transient openings for larger objects.

Therefore, we have developed a theoretical description of the diffusion of nonbinding proteins in
a dynamical polymer chain network. 

Theroetical descriptions of dynamical
polymer chain networks have already been given \cite{zhang}; however, they did
not incorporate realistic structures of  chromatin fibers. In the
following, the chromatin fibers are approximated by flexible polymer
chains. Proteins diffusing through the fiber network are referred to as
particles. We investigate in detail how the mean
square displacement (msd) of diffusionally transported particles of various
sizes depends on the three dimensional geometry of static and dynamical
networks of chromatin fibers and their density in the cell nucleus. The
results from simulations in both network types are compared. We then prove
by statistical analysis of the maximum mean square displacement of a
particle that a dynamical chain network facilitates the diffusional transport
of large particles. 

The chromatin network in the cell nucleus during the
interphase is modeled on a three-dimensional lattice applying
a simplified version of the bond fluctuation method \cite{carmesin}, the
single site model, in combination with a Metropolis Monte Carlo procedure
\cite{metropolis}. 
This minimizes computational time and effort.
The bond fluctuation model (BFM) was introduced as a new
effective algorithm for the dynamics of polymers by Carmesin et
al. \cite{carmesin} and provides a very effective means to simulate the
dynamic behaviour of, e.g., dense polymer melts \cite{sommer1, sommer2}. In
section 2 the lattice for chain construction and particle diffusion is
presented. A brief overview of the particles of different sizes is given,
reflecting the diffusing proteins in the cell nucleus. The simulation of chain
dynamics of one single chain is described. Afterwards, the simulation of chain
dynamics of dense chain networks, a combination of BFM and MC, is briefly
explained. In section 3 the translational diffusion of the chains is
observed. It is verified that the dense networks investigated here display a
semidilute behaviour and exhibit Rouse dynamics. Finally, in section 4
the mean square displacements of particles of various sizes in static and
dynamical networks are compared. It is shown that the diffusion of large
particles is facilitated by a dynamical chain network and how data of the
dynamical chain network can be reconstructed from data of the static chain
network. Moreover, it is demonstrated that an anomalous behaviour of the mean
square displacement of the particle on short times is caused by a trapping
process of the particle in the crowded environment of polymer chains.
Section 5 summarizes the obtained results.


\begin{figure}[ht] \centerline{
\includegraphics[scale=.38]{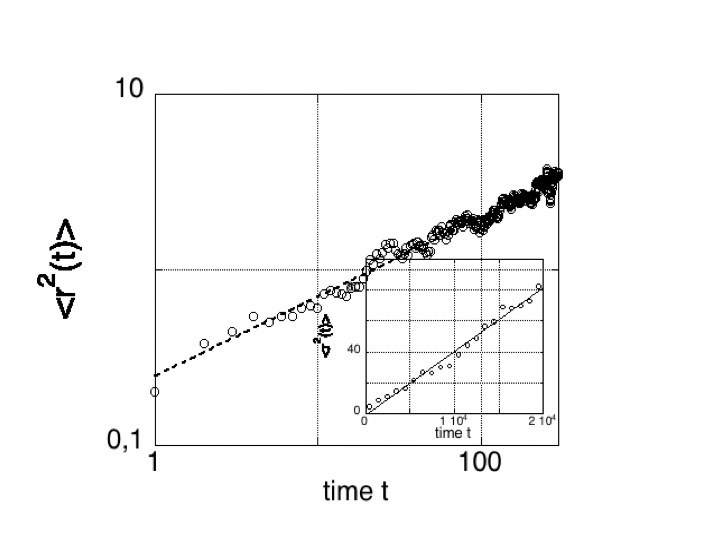} }
\caption{\label{rousemsd} Mean square displacement (msd) of one chain
monomer. Data is obtained by averaging over 50 chain conformations. Msd
behaviour for short time scales sketched on a logarithmic scale. Dashed line:
power law fit, anomaly parameter $\alpha=0.45$. Inset: msd behaviour for long
time scales. Straight line: linear fit.}
\end{figure}

\section{The model system} The system is contained in a $100\times100\times100$
cubic lattice. To prevent boundary effects due to the lattice walls, periodic
boundary conditions are applied for both the particle motion and the chain
construction.

\subsection{Simulation of particle movement} 
In general studies tracer particles with masses
of 27 to 282 kD were used to study diffusional transport \cite{banks}. In our
model system we use three
different particle
sizes: one occupied lattice site (small), a $2^3$ cube (medium), and a $3^3$
cube (large). The movement of the particle is modeled by a random
walk. Particles are allowed to visit only unoccupied lattice sites. If a
particle collides with a chain, it is reflected to the last visited lattice
site. This counts as a time-step, even though the move was unsuccessful.
The initial position is sampled randomly among nonoccupied lattice
sites. Then, a number of random trajectories (typically of the order
of $10^3$) are simulated to deliver the ensemble averages. 

\subsection{Chain dynamics}
\label{chaindynamics} The chromatin fibers are modeled as chains consisting of
monomers connected by freely jointed segments. Monomers are represented by 
occupied lattice
sites. No lattice site is allowed to be occupied more than once, otherwise
there exists no pair interaction between monomers. This corresponds to a
real chain with excluded volume effect in a good (more precisely: athermal)
solvent. Allowed
bonds between two adjacent monomers are taken from the set of all component
permutations and sign inversions $P$ of two bond vectors, $P(1,0,0)\cup
P(1,1,0)$, inducing a bond length of $1$ or $\sqrt 2$. Chain movements
are simulated with a combined version of the BFM and MC, which is discussed
in detail in \cite{wedemeier1}.

\begin{figure}[ht] \centerline{
\includegraphics[scale=.4]{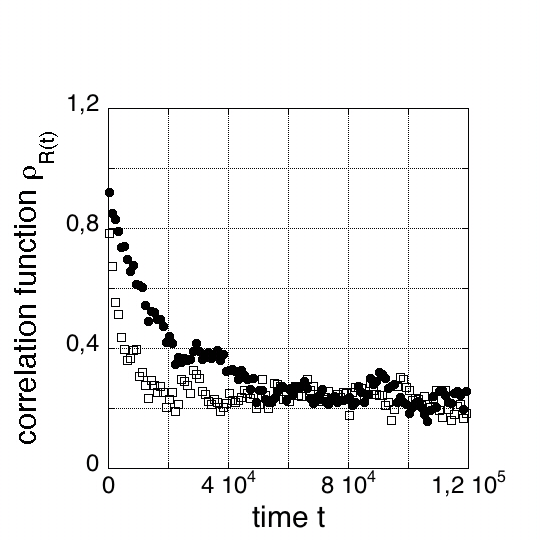} }
\caption{\label{rousetime} Correlation function $\rho_{R}$ as a function of
time $t$ for different chain lengths $N$. Squares: $N=50$, circles: $N=100$.}
\end{figure}

\subsubsection{Single chain dynamics} 
Although this paper is dealing with the diffusion of particles inside
a chain network, it is instructive to first analyze the dynamics 
of single chains, in order to obtain the time scales which are 
required to simulate ensemble averages of the network. 
Thus, we investigated the dynamical behaviour of single chains of
different lengths $N$, $N\in\{10,30,50,100\}$. 
Initially, a straight line of $N$ connected neighbouring lattice 
sites is placed into the lattice and relaxed into its equilibrium
conformation. During one time step (sweep) of the MC procedure, the $N$ monomers 
of the chain sequentially undergo a trial-move.
If this move hits an occupied lattice site or violates
the bond length restriction, the chain monomer is set back to its previous
position and the next monomer is addressed. 

\subsubsection{Chain network} The combined  BFM and MC
procedure is applied to an initial conformation of chains folded into cubes. 
See \cite{wedemeier1} for a detailed
description of the cube construction. The relaxed conformation then serves as 
the initial conformation for investigations of the dynamical network.
Similarly to the
case for single chains, all $k\times N$ monomers of the $k$ chains, each of
length $N$, are touched during a single time step (sweep) of the MC procedure.


\section{Results - Chain dynamics}
\label{sec:chaindynamics}
\subsection{Single chain} The center of mass movement
$\langle(x_{c.m.}(t)-x_{c.m.}(0))^2\rangle$ over time $t$ is investigated for
different single chains of chain length $N$, $N\in\{10, 30, 50\}$.
$x_{c.m.}(t)$ is the center of mass position vector of one chain at time
$t$. Results are the average of 50 independent runs.
The translational diffusion coefficient $D_{trans}$ of the chain is 
obtained from the mean squared center of mass displacement : 
\begin{equation} 
6tD_{trans}=\langle(x_{c.m.}(t)-x_{c.m.}(0))^2)\rangle.
\end{equation} 
First it was verified that the translational diffusion coefficient $D_{trans}$ 
decreases roughly inversely proportional to the chain length $N$,
\begin{equation} 
D_{trans}\sim \frac{1}{N}\;.
\label{scalingsingle}
\end{equation}
This is to be expected, because our simulations do not consider
effects of hydrodynamic correlations (immobile solvent approximation)
which should lead to the observed $N^{-1}$ (Rouse) scaling \cite{rouse} 
of the diffusion coefficient. It is generally assumed that these
hydrodynamic effects are playing a less important role inside dense
polymeric systems, because they are screened on rather short
length scales. Their omission in the present work is therefore 
justified, considering other, more severe approximations which come
along with our coarse grained model.

The mean square displacement of a single chain monomer is linear in time 
on long time scales (Fig. \ref{rousemsd}, inset)
\begin{equation} 
\langle r^2(t) \rangle \sim t, \; t\rightarrow \infty.
\end{equation} 
On short time scales, however, an anomalous behaviour of that quantity
is found,
\begin{equation} 
\langle r^2(t) \rangle \sim t^\alpha, \; t\leq 300
\end{equation} 
with an anomaly parameter $\alpha$ of 0.45 (Fig. \ref{rousemsd}).
 It can be shown that in case of the ideal chain approximation,
the anomaly parameter of such a monomer should yield approximately
0.5 \cite{grossberg} on short time scales. This is a result of the 
restricted movement of the monomer due to its bonds with other chain
monomers. The observed anomaly parameter of $\alpha = 0.45$ is 
a little smaller, which we attribute to additional excluded 
volume restrictions of our real (= non-ideal) chain model. 
On long time scales, of course, each monomer has to follow
the movement of the chain center of mass and hence displays 
the same diffusion behavior.


For the simulation of dynamical dense chain networks
it is important to know the relaxation times of the
chains involved in the simulation, because proper ensemble
averages require simulation times much larger
than these relaxation times. The slowest relaxation
mode of an ideal linear chain is the relaxation of its end-to-end vector $R$,
also called Rouse time $\tau_{R}$. It is obtained through the decay of 
the time autocorrelation function
\begin{equation} 
\rho_{R}(t)=\frac{\langle R(0)\times R(t)\rangle}{\langle R^2 \rangle}\;.
\end{equation} 
To extract the decay time, it is sufficient to assume a single mode
exponential decay of the form
\begin{equation} 
\rho_{R}(t)=\exp \left(-\frac{t}{\tau_{R}} \right).
\end{equation} 
In this way, the Rouse time $\tau_{R}$ of a single chain with
length $N=50$ was found as $\tau_{R}\approx 2\times 10^4$, and for a chain of
twice the length, $N=100$, the result $\tau_{R}\approx 8\times 10^4$ was obtained
(Fig. \ref{rousetime}). The Rouse time of an ideal chain can be shown to scale with 
the square of the number of monomers \cite{rubinstein},
\begin{equation} 
\tau_{R} \sim N^2\;,
\end{equation} 
and the same behavior is approximately true for real chains.
Thus, an important rule of thumb for simulating systems 
of different chain lengths is that the relaxation 
time of the system scales with the square of the chain length.  

\begin{figure}[ht] \centerline{
\includegraphics[scale=.4]{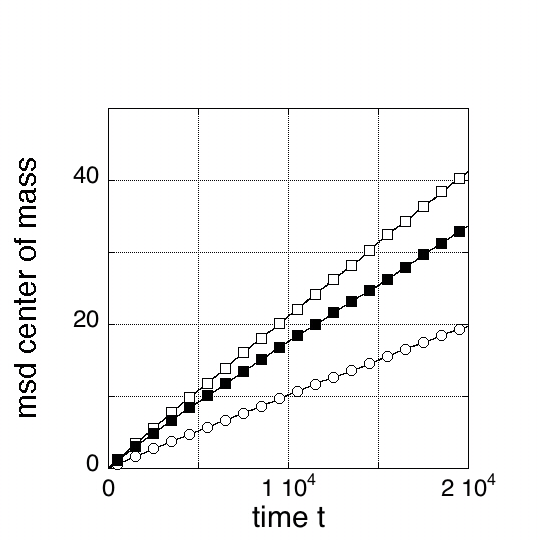} }
\caption{Center of mass vs. time $t$ for different chain lengths $N$ in a
dynamical chain network with different volume fractions. Squares: $N=50$,
(white: $D_{trans}=0.0021$, black: $D_{trans}=0.0017$)
circles: $N=100$ ($D_{trans}=0.001$). Blank symbols: volume fraction of $6.4\%$.  Solid
symbols: volume fraction of $12.5\%$.  Straight line: linear fit.
\label{cmnetwork} }
\end{figure}
\subsection{Chain networks} 

In order to characterize the chain network, it is required
to verify in which density regime the system exists. Our chain model
is consistent with a real chain inside an athermal solvent, and hence the
fully relaxed end-to-end distance of a single chain should yield an 
average of $R = d\,N^{\nu}$, where $d \approx 1.2$ is the average bond 
length and $\nu$ the universal scaling exponent, $\nu \approx 0.588$. 
This yields the overlap concentration of the system in athermal 
solvent \cite{rubinstein}
\begin{equation}
\phi^* \approx d^{6\nu - 3} N^{1 - 3\nu}\,.
\end{equation}
Hence, we obtain $\phi^*(N=100) \approx 0.03$ and $\phi^*(N=50) \approx 0.06$.
Volume fractions below this overlap concentration deliver dilute solutions
(in which single chains remain unaffected by other chains), while 
volume fractions above that threshold deliver semi-dilute solutions.
The volume fractions in our system were between $0.064$ and $0.125$, 
above the overlap concentration, hence the system is in the  
semi-dilute regime. The $N^{-1}$ Rouse scaling of the diffusion
coefficients remains valid under these conditions, but the pre-factors 
are affected by inter-chain interactions, i.e.\ the chain diffusion 
coefficient differs from the single chain case discussed above. 


The center of mass movement of a chain inside  dense networks
with different geometric volumes $\sigma$, $\sigma \in \{6.4\%, 8\%, 10\%,
12.5\%\}$ and different chain lengths $N$, $N \in \{50, 100\}$ was monitored
over the time $t$. The data was averaged over 10 simulation runs.
The denser the chain network, the slower is the movement of the chains and the
smaller is the translational diffusion coefficient of the chains
(Fig. \ref{cmnetwork}). For a chain network of
6.4\% volume and chains with length $N=50$ a translational diffusion
coefficient of $D_{trans}=0.002$ is obtained, whereas chains of length $N=100$
with the same volume yield a translational diffusion coefficient of
$D_{trans}=0.00098$ (Fig. \ref{cmnetwork}). Hence, the Rouse scaling 
law as for single chains (see equation (\ref{scalingsingle})) is 
reproduced, though with a different pre-factor,
\begin{equation} 
D_{trans-network}\sim \frac{1}{N}\,.
\end{equation}

\begin{figure}[ht] \centerline{ {\bf a}
\includegraphics[scale=.4]{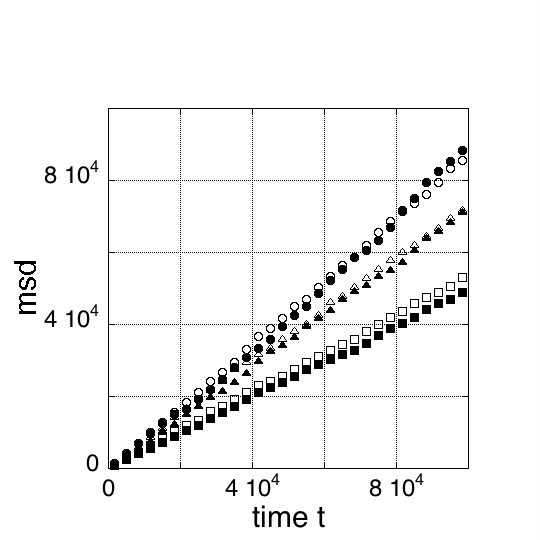} } \centerline{
{\bf b}
\includegraphics[scale=.4]{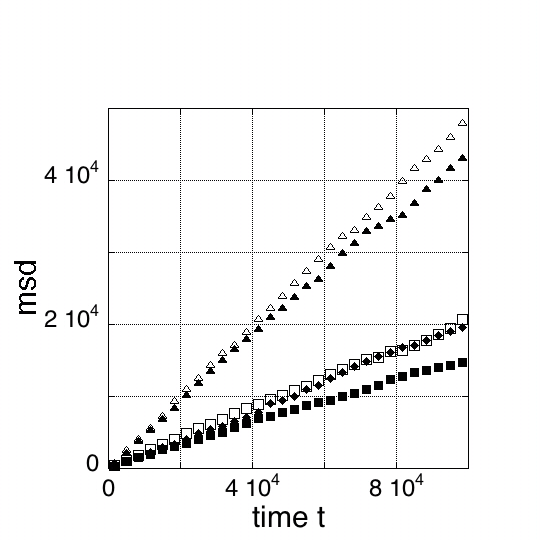} }
\caption{\label{msd} Mean square displacement vs. time $t$ for different 
particle sizes and volume fractions both in static and dynamical
networks. White symbols: dynamical chain network, black symbols: static chain network.
Chain length: $N=50$. Circles: $1\times1\times1$ particle (small), 
triangles: $2\times 2\times 2$ particle (medium-sized),
squares: $3\times 3\times 3$ particle (large),
diamonds: $3\times 3\times 3$ particle in a static network with weighting factors.
Upper panel: 6.4\% chain volume. Circles: $D_{static}=D_{dynamical}=0.87$.
Triangles: $D_{static}=D_{dynamical}=0.72$. 
Squares: $D_{static}=0.49$, $D_{dynamical}=0.53$.
Lower panel: 12.5\% chain volume.
Triangles: $D_{static}=0.44$, $D_{dynamical}=0.48$.
Squares: $D_{static}=0.15$, $D_{dynamical}=0.21$.
Diamonds: $D_{weightingfactor}=0.21$
} 
\end{figure}

\section{Particle diffusion in dynamical chain networks}

In the following the occupation volume of the
lattice is either computed as the geometric volume or the effective
volume. The geometric volume is defined as the total number of occupied
lattice sites. The effective volume is defined as the space which is blocked
to a particle of given size. Hence, the effective volume of
the chains on the lattice depends on the particle size and the chain
conformation. For a particle consisting of one occupied lattice site, 
the geometric volume of the chain equals its effective volume.

\subsection{Coupled dynamics of particle and chain diffusion} 
During one time step of the MC procedure, an entire sweep over the
chains was carried out, as described in section \ref{chaindynamics},
and additionally a single move of the walker. This setup corresponds
to the choice of identical diffusion coefficients of the walker (protein)
and a single (non-connected) statistical chain monomer. In other words,
a chain monomer and a protein were assumed to be of roughly the same
size. This does not impose any fundamental restrictions, because more 
general situations could be obtained after a corresponding rescaling 
of the diffusion coefficients. In order to increase the efficiency,
fifty walkers were moving through the network at the same time.
They were independent of each other, i.e.\ there was no walker-walker
interaction, to simulate the ensemble average of single walker systems.
The initial positions of the walkers were randomly sampled inside
the lattice, avoiding occupied lattice sites. The simulations  
were run for $t_{MC}=10^5$ MC steps; after one
run was completed, a new set of 50 walkers was initialized. Ten
runs were averaged in this way. Since the simulation time
of each single run did exceed the Rouse relaxation times of the
chains $t_{MC} > \tau_{R}$, it was assured that each run was featuring
a statistically independent set of chain conformations. 
The occpuation volume was varied between $6.4\%$ and $12.5\%$.

\subsection{Physical time scales} 
Our simulations yield values for the diffusion
coefficients $D$ of the walker that are scaled  with its diffusion coefficient 
inside an empty lattice, $D_0$. Consequently, if the free diffusion 
coefficient $D_{\rm 0, exp}$ of a protein is known experimentally, it may 
be multiplied with the diffusion coefficient of our simulation to yield 
the approximate diffusion coefficient of the protein inside the network.

Alternately, the Einstein-Stokes relation may be applied to estimate
real time diffusion coefficients. In our coarse grained model, one
lattice site approximately corresponds to one Kuhn length of the DNA,
which amounts to $b \approx 100 nm$. Let us assume that we know the
physical time required for the protein to diffuse a distance equal
to $b$, and call it $\tau_0$. This corresponds to a
single time step in our random walk simulation, while $b$ stands for
the step-size. Hence our diffusion law may be interpreted as
\begin{equation} 
\left\langle \left( \frac{r(t)}{b} \right)^2\right\rangle= 
6 D \left(\frac{t}{\tau_0}\right)\;,
\end{equation}
where $r(t)$ is a real physical distance, $t$ a physical time and $D$
the diffusion coefficient found in the simulation. In order to find
an approximation for $\tau_0$, we assume the protein to diffuse freely
on the length scale of one Kuhn-length (i.e.\ one lattice site)
and employ the Einstein-Stokes relation
\begin{equation} 
b^2 = 6D_{\rm ES}\; \tau_0\;,
\end{equation}
where
\begin{equation} 
D_{\rm ES}=\frac{k_BT}{6\pi\eta  R}\;.
\label{einstein_stokes}
\end{equation} 
Here, $k_B$ is the Boltzmann constant, $T$ the absolute temperature,
$\eta$ the viscosity of the solvent and R the hydrodynamic radius of
the protein. As an example, let us consider a (quite large) protein with a
hydrodynamic radius of $R = 15 nm$,  $T=300 K$ and $\eta=10^{-3} N\cdot s\cdot
m^{-2}$. With $k_B=1.3\cdot 10^{-23} \frac{J}{K}$ we then obtain
$D_{\rm ES} \approx 1.4\cdot 10^{-11} \frac{m^2}{s}$. This yields a
fundamental time unit of $\tau_0 \approx 1.2\cdot 10^{-4} s$. 
Hence, a typical simulation run ($10^5$ time-steps) would correspond 
to roughly $10$ seconds of physical time and cover a distance of the 
order of $10 \mu m$.

\subsection{Comparison of particle diffusion in static and dynamical polymer
networks} 
Fig. \ref{msd} a displays the dependence of the mean square
displacement of particles on the time $t$ both in
static (black symbols) and dynamical (white symbols) networks with a geometric chain
volume of 6.4\%. For the
smallest particle the mean square displacement shows the same behaviour
independent of the network dynamics.
Fig. \ref{msd} b displays the
dependence of the mean square displacement of  particles on the time
$t$ both in static (black symbols) and dynamical (white symbols) networks with a geometric chain
volume of 12.5\%. 

The
mean square displacement of the largest particle in a dynamical chain network
with an occupation volume of 6.4\% (Fig. \ref{msd} a, squares) is higher than in a corresponding static
chain network. The diffusion coefficients and their difference concerning both
network types are similar to those obtained from simulations with the
medium-sized particle in a chain network with an occupation volume of 12.5\%
(see Fig. \ref{msd} b, triangles) due to the same effective chain volume of
$\sigma_{eff}\approx 52\%$ in both systems.

\begin{figure}[ht!]  \centerline{
\includegraphics[scale=.4]{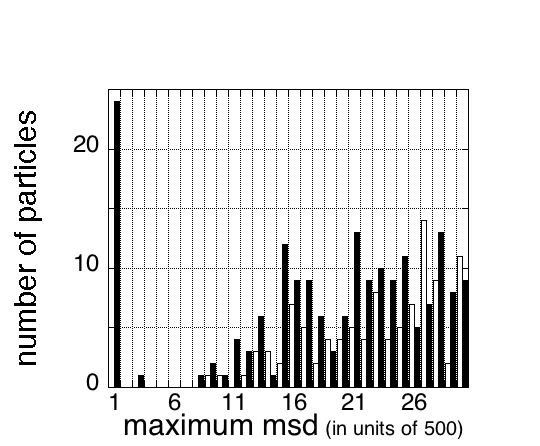} }
\caption{\label{bar3x3x3} Maximum mean square displacment of a
particle. Number of particles: 500.  Black bars: static network, white bars:
dynamical network.  $3\times3\times3$ particle (large), chain volume: 12.5\%.
For a better overview only maximum msd values less than 15000 lattice units
are shown.}
\end{figure}

The mean square displacement of
the largest particle in a dynamical chain network with an occupation volume of
12.5\% (Fig. \ref{msd} b, squares) is on long time scales significantly higher than in a corresponding
static chain network. This does not imply that each single walker is moving 
faster through the dynamic network. Fig. \ref{bar3x3x3} shows that
around $5\%$ of the particles (24 out of 500) did not reach a maximum mean 
square displacement beyond 500 (lattice units)$^2$
(black bars), which means that they remained trapped within
rather small pockets inside the network. In a corresponding dynamical network all
particles reached a maximum mean square displacement larger than 500 (lattice units)$^2$
(white bars), which naturally results in a larger
averaged value for the diffusion coefficient. This finding indicates that 
most of the walkers did hardly modify their diffusion
inside the dynamic network. A minority of walkers, however, which would
remain trapped in the static case, becomes released in the dynamic case
and in this way modify the average diffusion coefficient.  
Therefore, the difference in the diffusion
coefficients is a direct consequence of the averaging process during the
computation of the mean square displacement curve.

In order to verify this finding 
in a quantitative way, the trajectories of the particles in the static chain
network were multiplied with a weighting factor determined by the ratio of the
both bar lengths (dynamical/nondynamical). In other words:
A walker which was moving over a short distance
got a lower statistical weight than a particle which was moving over a longer
distance. With this approach, we succeeded to reconstruct the data of the
dynamical chain network from the data of the static chain network (see
Fig. \ref{msd} b, diamonds).

It is important to mention that the walker movement is not only
influenced by the fluctuations of the chain network but also by the center
of mass diffusion of the chains. A pure movement of the walker due to the 
center of mass diffusion of the chains can be neglected:
the difference of 0.06 in the diffusion coefficients of the static and dynamical
chain network with 12.5\% volume occupation (Fig. \ref{msd} b) for the largest particle
is much larger 
than the translational diffusion coefficient of the chains $D_{trans}=0.0017$ 
(Fig. \ref{cmnetwork}) indicating that the movement of the walker is mainly 
influenced by the chain fluctuations.

\begin{figure}[ht] \centerline{ {\bf a }
\includegraphics[scale=.4]{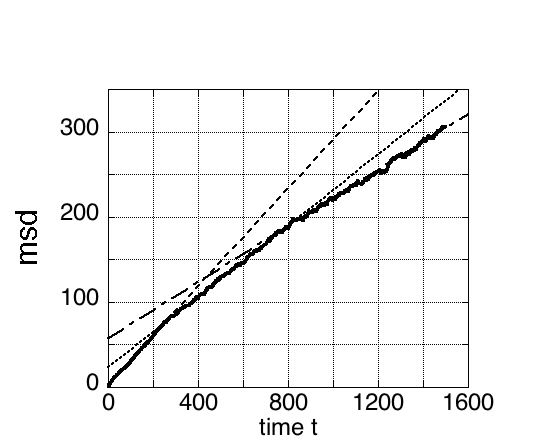} } \centerline{ {\bf b }
\includegraphics[scale=.4]{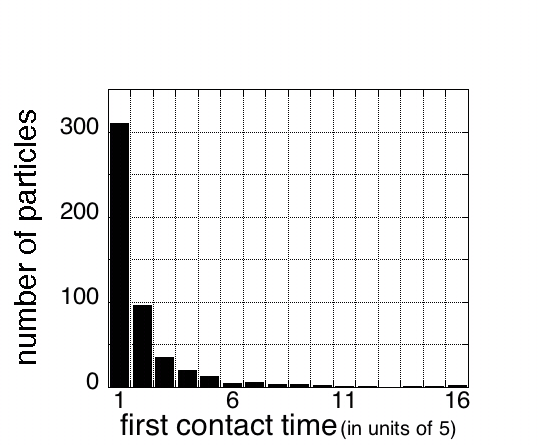} }
\centerline{ {\bf c}
\includegraphics[scale=.4]{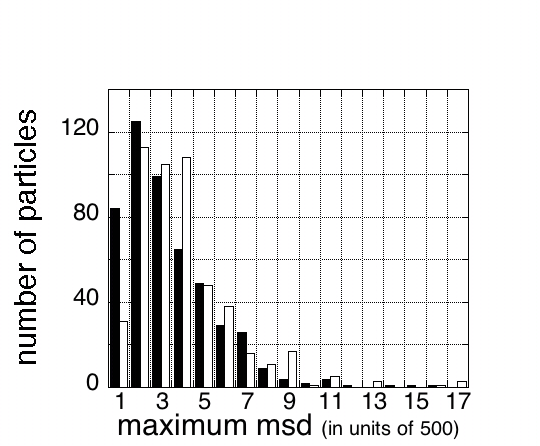} }
\caption{\label{fct} $t\leq 5000$. Upper panel: Mean square displacement
vs. time $t$ on short time scales. Static chain volume: 12.5\%,
$3\times3\times3$ particle (large). Middle panel: First contact time of the
particle with the static chain network. Number of particles: 500. Chain
volume: 12.5\%, $3\times3\times 3$ particle (large). Lower panel: maximum mean
square displacement. Number of particles: 500. Chain volume: 12.5\%, $3\times
3\times 3$ particle (large).}
\end{figure}

\subsection{Subdiffusion on short time scales} Simulations on short time
scales with the number of MC steps $t_{MC}\leq 5000$ were carried out to
investigate the subdiffusive behaviour of the mean square displacement of
diffusing particles for short times. 

In a static chain network with a chain
volume of $12.5\%$ the mean square displacement of the largest particle displayed
subdiffusive behaviour with an anomaly parameter $\alpha=0.81$ (Fig. \ref{fct}
a), in a corresponding dynamical chain network $\alpha=0.83$ was observed, 
not significantly different when considering the statistical fluctuations.
In Fig. \ref{fct}, three different subregimes are observed
(dashed lines): starting with a diffusion coefficient of $D=0.29$, the
particles become slower after almost 400 MC steps. After 800 MC steps the
diffusion coefficient decreases to $D=0.16$. One might suspect that the first
collision time of the particle could be responsible for this decrease in the
diffusion coefficient: the particle diffuses freely until it encounters the
chain network. However, as displayed in Fig. \ref{fct} b, the majority
of particles are colliding with the network within the first five time steps.
Instead, the decreasing diffusion coefficient
is caused by the increasing number of trapped particles: $16.8\%$ of the 
investigated particles (84 out of 500) did not reach a maximum mean square 
displacement larger than 500. 

Thus, subdiffusion on short time scales is not an
effect of first collision times of the particle with the chain
network. Instead, subdiffusion results from the fact that on short time scales
particles are trapped in a crowded environment,  hence do not contribute to the
large distances of the mean square displacement on longer time scales.

\section{Discussion} 
The present work was intended to shed some light on 
 the diffusional transport of proteins in the dynamical 
chromatin network in the cell nucleus. This process is described with a lattice
model of the nucleus, the chromatin fibers and a random walk of the
proteins. As shown in earlier work \cite{wedemeier1, wedemeier2} the lattice
model has - compared to a corresponding continuum model - the advantage of
being roughly two orders of magnitude faster due to the finite number of
states. 

To create crowded environments of chromatin fibers in the cell
nucleus, a self-avoiding random walk chain model with excluded volume 
 was used, and chain dynamics was implemented
through the BFM and MC. The findings in Section \ref{sec:chaindynamics}
are hardly surprising to those who are familiar with polymer dynamics.
The simulation of polymeric systems however, is not trivial due
to the different  time scales involeved, and one 
has to verify that one does in fact obtain proper
ensemble averages. Thus, we verified the known scaling laws and
typical relaxation times.

It may be argued that the chain lengths used in this work might be too small
to represent a chromatin fiber inside a cell nucleus. In reality, the latter
would usually cover ten thousands of Kuhn monomers and create a chain 
network far inside the concentrated regime (and not, as in our case, in
the semi-dilute regime). Moreover, such a chromosome would not exhibit
any significant center of mass motion because it remains confined inside
its territory. In reality, these differences are of little relevance here.
The walker does, at any stage of the simulation, only ``see'' a small
subsection of the chain. Its dynamics is therefore influenced by the
fluctuations of local chain sections, and hence it is irrelevant how
far the chain does extend beyond the walker's limited horizon. In the
semi-dilute regime, remote sections of the same chain are already
uncorrelated, so that a single long chain might as well be replaced with 
several independent shorter chains. Additionally, as a
result of the Rouse-scaling, the chain center of mass diffusion was
roughly two orders of magnitude slower than the walker's diffusion
and hence it did contribute to its motion only on the level of
a few percent.

In several systematic simulations the mean
square displacements of diffusing particles of various sizes were computed
both in static and dynamical polymer networks. By comparing the results of the
two network types we could show that chain diffusion does not alter the
diffusion process of small particles. However, the polymer network dynamics
begins to influence the diffusional transport process of particles once the
effective volume of the chain network is reaching 50\%
(Fig. \ref{msd}). Hence, chain dynamics facilitates the transport of large
particles. It was demonstrated that the difference in the diffusion
coefficients in both network types is a direct consequence of the averaging
process during the computation of the mean square displacement. Hence
we were able to reproduce the results of the dynamical network after 
statistically reweighting the single walkers contributions to the averages
obtained within the static network. It might be interesting to investigate
whether such a procedure could be formalized, so that the simulation
of dynamical networks could be replaced with static network simulations,
augmented with a corresponding weight function for each trajectory.

In further simulations anomaly on short time scales is found for the msd of a
particle in a dynamical fiber network. We could show that this kind of
subdiffusion does not result from the first collision time of the particle but
is caused by a relatively high number of trapped particles in the crowded
environment of the fiber network (Fig. \ref{fct}).


In recent years, chromatin-binding proteins have been extensively
characterized due to their
functional significance and their dynamics in the living cell has been studied by in
vivo microscopy techniques \cite{houtsmueller, lippincott}. However, it is not
yet clear how proteins bind to the chromatin fibers and displace themselves
within the dynamical chromatin network \cite{richter, berg1}.

We believe
that using the model of diffusional transport of proteins in a dynamical fiber
network presented here, it is feasible to clarify these questions with an
additional implementation of a binding process and with the help of further
systematic simulations. This is ongoing work.


\section{Acknowledgement}
One of the authors (A.W.) thanks two of the authors (H.M. and C.W.) for their hospitality during a research stay at the Department of Physics at the Xiamen University. One of the authors (A.W.) was supported by a scholarship from BIOMS (Center for Modeling and Simulation in the Biosciences). In addition, A.W. thanks H. Soller for a fruitful discussion concerning physical time scales.

\end{document}